\documentclass[10pt,letterpaper,twocolumn]{article} 

\newcommand{\Veff}{$V_\textrm{eff}$}
\newcommand{\wres}{$\omega_\textrm{res}$}
\newcommand{\lambdares}{$\lambda_\textrm{res}$}

\usepackage{ol2nocopyright}
\usepackage{amsmath}
\usepackage{epstopdf} 

\usepackage{color}

\begin{document}

\twocolumn[ 

\title{Optical design of split-beam photonic crystal nanocavities}

\author{Aaron C.\ Hryciw$^{1,*}$ and Paul E.\ Barclay$^{(1,2)}$}

\address{
$^1$National Institute for Nanotechnology, 11421 Saskatchewan Drive, Edmonton, Alberta, Canada, T6G 2M9\\
$^2$Institute for Quantum Information Science, University of Calgary, 2500 University Drive NW, Calgary, Alberta, Canada, T2N 1N4\\
$^*$Corresponding author: Aaron.Hryciw@nrc-cnrc.gc.ca
}

\begin{abstract}
We design high quality factor photonic crystal nanobeam cavities formed by two mechanically isolated cantilevers.  These ``split-beam'' cavities have a physical gap at the center, allowing mechanical excitations of one or both of the cavity halves. They are designed by analyzing the optical band structures and mode profiles of waveguides perforated by elliptical holes and rectangular gaps, and are predicted to support optical resonances with quality factors exceeding 10$^6$ at wavelengths of $\sim$1.6 $\mu$m.
\end{abstract}


 ] 

\noindent 

Photonic crystal (PC) nanobeam cavities \cite{ref:notomi2008uqn} are currently the subject of intense research interest, as they possess a suite of attractive characteristics:  high optical quality factor ($Q$), wavelength-scale effective mode volume (\Veff), low effective mass, and small physical footprint. Integrated within planar photonic waveguide circuitry, PC nanobeam cavities provide an ideal architecture for sensing via cavity optomechanics \cite{ref:eichenfield2009pns,ref:eichenfield2010oc,ref:deotare2012aor}.   Many sensing applications are poised to benefit from orders-of-magnitude improvements in sensitivity enabled by on-chip cavity optomechanics, including measurement of mechanical resonators \cite{ref:gavartin2012hoc}, inertial sensing \cite{ref:krause2012ahr}, novel platforms for chip-based AFM \cite{ref:srinivasan2011oti}, and detection of torsional motion for magnetometery \cite{ref:davis2010nrt,ref:kim2012nto}. To maximize the sensitivity of these measurements, devices whose central cavity region is not rigid \cite{ref:li2009bap}, and can be easily deformed by external forces \cite{ref:eichenfield2009pns, ref:srinivasan2011oti, ref:wu2013top} are desirable.  To this end, here we describe  the optical design of a PC nanobeam cavity containing a complete cut through the cavity center, yielding an optical cavity formed by two mechanically uncoupled nano-cantilevers.  We show that by careful consideration of the anomalous ordering of the instantaneous band-edge modes at the center of the cavity, a high-$Q$ cavity can be realized in this disconnected structure.

\begin{figure}[t]
\centerline{\includegraphics[width=8.3cm]{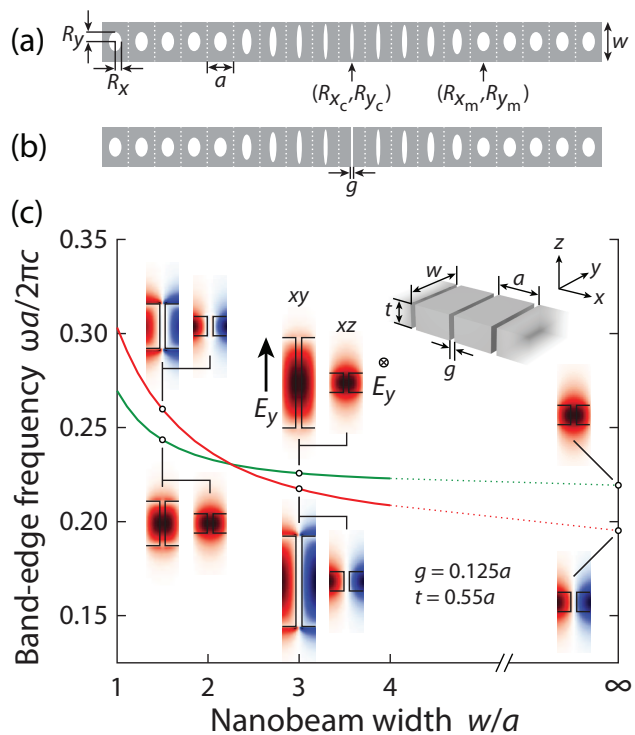}}
\caption{\label{gapWGmodes}(a) Donor-mode PC nanobeam cavity with elliptical holes. (b) ``Split-beam'' PC cavity with a central rectangular gap.  (c) ``Gap unit cell'' band-edge frequencies and electric field profiles for TE-like ($y$-odd, $z$-even) modes vs.\ nanobeam width.}
\end{figure}

We begin with the deterministic cavity design approach of Quan \emph{et al.} \cite{ref:quan2011ddw,ref:quan2010pcn}, the essence of which is to taper the unit cell geometry of a host PC waveguide  to effect a linear change in the mirror strength $\gamma$, yielding an approximately Gaussian cavity field envelope.  Near the band edge of a PC waveguide, the  wavevector may be approximated by $k_x=(1+i\gamma)\frac{\pi}{a}$, with
\begin{equation}
\label{gammaeq}
\gamma=\left[\left(\frac{\omega_2-\omega_1}{\omega_2+\omega_1}\right)^2-\left(\frac{\omega_\textrm{res}-\omega_0}{\omega_0}\right)^2\right]^{\frac{1}{2}}
\end{equation}
where $\omega_1$  and $\omega_2$ are the lower and upper (respectively) band-edge frequencies defining the optical bandgap of a periodic waveguide, $\omega_0$ is the mid-gap frequency, and \wres~is the cavity resonance frequency \cite{ref:quan2010pcn}.  A high-$Q$ cavity may be achieved by choosing a host waveguide such that $\gamma$ at the cavity center is maximized for \wres~(minimizing power leakage into the waveguide), and tapering $\gamma$ away from the cavity center sufficiently adiabatically to limit scattering into radiation modes. A generalized donor-mode cavity (field concentrated in the air holes) with elliptical holes is shown in Figure \ref{gapWGmodes}a, tapering from the central hole $(R_{x_\textrm{c}},R_{y_\textrm{c}})$ to the ``mirror'' (host waveguide) hole $(R_{x_\textrm{m}},R_{y_\textrm{m}})$ over six segments.

A plausible (albeit na\"{\i}ve) program to construct a PC nanobeam cavity with a central gap is to start with a high-$Q$ design as explained above and simply cut the nanobeam at the cavity center, yielding a ``split-beam'' cavity capable of supporting a variety of mechanical modes (e.g., axial, torsional, in- and out-of-plane).  The mechanical properties could then be tuned by adjusting the nanobeam supporting structure \cite{ref:wu2013top}.  A donor-mode cavity, whose field is concentrated in the air regions, allows a strong field overlap with the central gap, whose width may be modulated by mechanical excitations. Intuition suggests that to limit out-of-plane scattering, the central hole unit cell be replaced with a gap unit cell whose band-edge frequency and spatial field profile closely match those of the hole.

To evaluate this  design quantitatively, we begin by analyzing a PC waveguide composed of a periodic array of ``gap unit cells" formed by dielectric blocks with lattice constant $a$ (along $\hat{x}$) and thickness $t$, separated by gaps of width $g$ (centered at $x=0$).  Figure \ref{gapWGmodes}c shows the band-edge frequencies and field profiles ($E_y$) for the lowest-frequency  TE-like ($y$-odd, $z$-even) modes as the block width $w$ varies, calculated using a 3D frequency-domain eigensolver \cite{ref:johnson2001mpb} with a spatial discretization of 32 per period $a$.  A relative permittivity of $\epsilon = 12.11$ (e.g., silicon at a wavelength of 1550 nm) is used throughout. 

The crossing  in Fig.\ \ref{gapWGmodes}c of the gap PC waveguide band edges reveals an important property of this structure which must be understood to  design a high-$Q$ nanobeam cavity with a central gap.  In the 2D limit ($w/a\to \infty$), the ordering of the band-edge frequencies may be explained following the analysis of 1D PCs\cite{ref:joannopoulos2008pcm}:  the field of the lowest-frequency TE band edge (red line) is concentrated in the dielectric, yielding a higher effective index than the second TE band edge (green line), in which the field is concentrated in the air gap.    As $w/a$ changes, the dispersion of the ``dielectric'' mode is larger than that of the ``air'' mode, causing the frequency of these modes to cross. Unlike in a 1D PC, for $w/a <2.25$, the lowest-frequency TE band edge of a gap PC waveguide is an air mode.  This crossing behavior is not exhibited in PC nanobeam waveguides with a hole in place of the gap for the considered range of $w/a$. The difference in dispersion responsible for the crossing can be understood from perturbation theory of  moving boundaries \cite{ref:johnson2002ptm}:  as $w$ changes, the dielectric band edge is shifted more quickly than the air band edge because the dielectric band-edge field overlaps more strongly with the moving waveguide boundary.  In contrast, the air band-edge field is concentrated in the gap region, which undergoes no change in shape. This crossing is of crucial importance when designing split-beam cavities, as it governs which of the two gap unit cell band-edge modes should be matched with the band-edge \emph{air mode} (or \emph{dielectric mode} for an acceptor cavity) of the removed central hole.

\begin{figure}[t]
\centerline{\includegraphics[width=8.3cm]{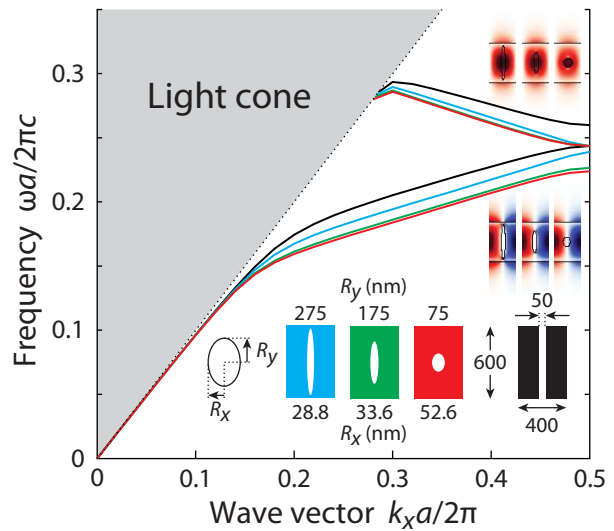}}
\caption{\label{modematching}Band structures of a ``gap unit cell'' and three elliptical holes unit cells with $\omega_{2_\textrm{ellipse}}=\omega_{1_\textrm{gap}}$.  Unit cell geometries are shown in the inset (all units in nm), with $t=220$ nm.  Field profiles of the dielectric (lower) and air (upper) modes for the elliptical holes are shown next to their associated bands.}
\end{figure}

To illustrate this design paradigm, we investigate three cavities with air-mode resonance wavelengths around 1.6 $\mu$m, suitable for fabrication using silicon-on-insulator (SOI) wafers.  For a nanobeam with thickness $t=220$ nm, width $w=600$ nm, and periodicity $a=400$ nm, a gap of width $g=50$ nm corresponds to Fig.\ \ref{gapWGmodes}c with $w/a=1.5$:  the air mode is the lowest-energy TE-like mode.  In Fig.\ \ref{modematching}, we compare the band structure of this rectangular gap unit cell with that of three elliptical holes, each satisfying the matching condition that the air-mode band-edge frequencies of the holes and the gap are equal (\wres$a/2\pi c$=0.2435).   For $y$ semi-axes of $R_{y_\textrm{c}}=275$, 175, and 75 nm, this is satisfied using $x$ semi-axes of $R_{x_\textrm{c}}=28.8$, 33.6, and 52.6 nm, respectively.

\begin{figure}[tb]
\centerline{\includegraphics[width=8.3cm]{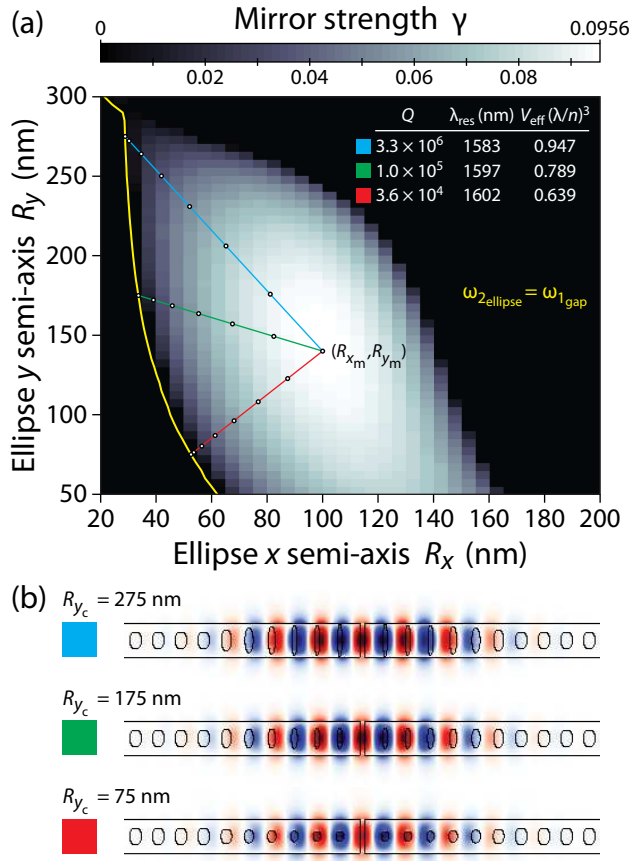}}
\caption{\label{cavdesign} (a) Mirror strength and design parameters for three split-beam cavities; $a$, $t$, and $w$, and the three $(R_{x_\textrm{c}},R_{y_\textrm{c}})$ values are as in Figure \ref{modematching}. The half-cavity tapered ellipse shapes are shown by open circles.  (b) Mode profiles ($E_y$, $xy$ plane) of the three cavities in (a).}
\end{figure}

With two parameters defining their shape---$R_{x_\textrm{c}}$ and $R_{y_\textrm{c}}$---elliptical holes satisfy the matching condition with a continuum of shapes, plotted with the yellow line in Fig.\ \ref{cavdesign}a.  The mirror strength $\gamma$ is calculated as a function of $(R_{x_\textrm{m}},R_{y_\textrm{m}})$ using \eqref{gammaeq}; as shown in Fig.\ \ref{cavdesign}a, the maximum occurs at $(R_{x_\textrm{m}},R_{y_\textrm{m}})=(100,140)$  nm.  To maintain a small nanobeam cavity footprint, we taper the ellipses between the central $(R_{x_\textrm{c}},R_{y_\textrm{c}})$ and mirror $(R_{x_\textrm{m}},R_{y_\textrm{m}})$ shapes over $N_\textrm{c}=7$ holes, capping each half-cavity with $N_\textrm{m}=8$  holes of shape $(R_{x_\textrm{m}},R_{y_\textrm{m}})$.  We achieve an approximately linear variation of $\gamma$ between the central and mirror holes by tapering both semi-axes quadratically:  $R_{x_j,y_j}=R_{x_\textrm{c},y_\textrm{c}}+(j/N_\textrm{c})^2(R_{x_\textrm{m},y_\textrm{m}}-R_{x_\textrm{c},y_\textrm{c}})$, for integer $j\in[-N_\textrm{c},N_\textrm{c}]$.  The hole shapes for the three cavities defined by the central holes in Fig.\ \ref{modematching} are plotted as circles in Fig.\ \ref{cavdesign}a.  The central ($j=0$) ellipse is then replaced by a 50-nm-wide rectangular gap to complete the design.  Note that, up to this point, only band-structure calculations have been used in the design \cite{ref:quan2011ddw}.

We assess the performance of these three cavities using finite-difference time-domain (FDTD) simulations \cite{ref:oskooi2010mff}; the field profiles for each are shown in Figure \ref{cavdesign}b, with $Q$,  \lambdares, and \Veff~(defined by the peak field strength) listed in the inset to Figure \ref{cavdesign}a.  The overall cavity $Q$ is dominated by scattering perpendicular to the nanobeam, split nearly evenly in the $y$ and $z$ directions; loss into guided modes ($x$ direction) is small due to the large $\gamma$ of the mirror segments.  While all three cavities satisfy the band-edge matching condition described above, significantly less scattering occurs as the fictitious ``central ellipse'' more closely resembles the cross-section of a rectangular gap (i.e., large $R_y$), achieving a $Q$ of $3.3\times10^6$ for the $R_{y_\textrm{c}}$=275 nm cavity.  As elucidated in \cite{ref:johnson2002atc}, scattering from the band-edge waveguide modes into lossy radiation modes generally increases as the taper between waveguide cross-sections becomes less gradual. In the case of the $R_{y_\textrm{c}}$=75 nm cavity, the poor adiabaticity between the gap and the neighboring hole degrades the $Q$ to $3.6\times10^4$. 

The band-edge matching condition yields ($R_{x_\textrm{c}},R_{y_\textrm{c}}$) values which are very close to optimal for a quadratic taper:  a free maximization of $Q$ with respect to $R_{x_\textrm{c}}$ using FDTD yields only a $\sim$20\% improvement in $Q$ for the $R_{y_\textrm{c}}$=275 and 175 nm cavities, with a change in $R_{x_\textrm{c}}$ of less than 10\%.  Alternately, $Q$ may be increased---at the expense of a larger \Veff~and cavity footprint---by increasing $N_\textrm{c}$ to yield a more adiabatic taper \cite{ref:quan2011ddw}.  From an experimental standpoint, Figure \ref{cavdesign}a illustrates the inherent trade-off in this scheme between $Q$ and practical fabricability:  pattern fidelity becomes more difficult for long, narrow elliptical holes due to minimum feature size limits in lithography and etching processes.

By studying the band structure of a nanobeam patterned with an array of gaps, we have demonstrated a technique for designing PC ``split-beam'' cavities with high $Q$ ($>$$10^6$) despite the existence of a major perturbation in the cavity region. These cavities support modes whose field is concentrated in the central air gap, and are expected to be well suited for sensing applications involving mechanical actuation of the nanocavity cantilevers.  This will be particularly useful for applications such as force sensing and torque magnetometry \cite{ref:kim2012nto}.

We thank M.\ H.\ Wu, J.\ D.\ Davis, and M.\ R.\ Freeman for helpful discussions.  This work was supported by NRC, CFI, $i$CORE/AITF, and NSERC.

\bibliographystyle{osa}  

\bibliography{library}

\end{document}